\documentclass[twocolumn,showpacs,preprintnumbers,amsmath,epsfig,apsrev,aps, superscriptaddress,prl]{revtex4}

\usepackage{graphicx}
\usepackage{graphics}
\usepackage{dcolumn}
\usepackage{bm}
\newcommand{\scr}{Sr$_3$Cr$_{2}$O$_8$}

\begin{document}
\preprint{}

\title{Asymmetric Thermal Lineshape Broadening in a Gapped 3-Dimensional Antiferromagnet - Evidence for Strong Correlations at Finite Temperature}

\author{D. L. Quintero-Castro}
\email[]{diana.quintero_castro@helmholtz-berlin.de}
\affiliation{Helmholtz-Zentrum Berlin f\"ur Materialien und Energie, D-14109 Berlin, Germany}
\author{B. Lake}
\affiliation{Helmholtz-Zentrum Berlin f\"ur Materialien und Energie, D-14109 Berlin, Germany}
\affiliation{Institut f\"ur Festk\"orperphysik, Technische Universit\"at Berlin, D-10623 Berlin, Germany}
\author{A. T. M. N. Islam}
\affiliation{Helmholtz-Zentrum Berlin f\"ur Materialien und Energie, D-14109 Berlin, Germany}
\author{E. M. Wheeler}
\affiliation{Helmholtz-Zentrum Berlin f\"ur Materialien und Energie, D-14109 Berlin, Germany}
\affiliation{Institut Laue Langevin, 6 rue Jules Horowitz, BP 156, F-38042, Grenoble Cedex 9, France}
\author{C. Balz}
\affiliation{Helmholtz-Zentrum Berlin f\"ur Materialien und Energie, D-14109 Berlin, Germany}
\affiliation{Institut f\"ur Festk\"orperphysik, Technische Universit\"at Berlin, D-10623 Berlin, Germany}
\author{M. M\aa{}nsson}
\affiliation{Neutron Scattering and Magnetism Group, Laboratory for Solid State Physics, ETH Z\"urich, CH-8093 Zürich, Switzerland}
\author{K. C. Rule}
\affiliation{Helmholtz-Zentrum Berlin f\"ur Materialien und Energie, D-14109 Berlin, Germany}
\author{S. Gvasaliya}
\affiliation{Neutron Scattering and Magnetism Group, Laboratory for Solid State Physics, ETH Z\"urich, CH-8093 Zürich, Switzerland}
\author{A. Zheludev}
\affiliation{Neutron Scattering and Magnetism Group, Laboratory for Solid State Physics, ETH Z\"urich, CH-8093 Zürich, Switzerland}

\date{\today}

\begin{abstract}
It is widely believed that magnetic excitations become increasingly incoherent as temperature is raised due to random collisions which limit their lifetime.  This picture is based on spin-wave calculations for gapless magnets in 2 and 3 dimensions and is observed experimentally as a symmetric Lorentzian broadening in energy. Here, we investigate a three-dimensional dimer antiferromagnet  and  find unexpectedly  that the broadening is asymmetric - indicating that far from thermal decoherence,  the excitations behave collectively like a strongly correlated gas. This result suggests that a temperature activated coherent state of quasi-particles is not confined to special cases like the  highly-dimerized spin-1/2  chain but is found generally in dimerized antiferromagnets of all dimensionalities and perhaps gapped magnets in general. 
\end{abstract}

\maketitle

In the conventional picture of temperature effects in magnetism, the excitations are long-lived at low temperatures and their lifetime decreases as temperature is increased. The accepted reason for this is that thermally activated excitations collide with each other limiting their lifetimes which results in a Lorentzian energy broadening of the lineshapes \cite{harris,Ty}. This model works well for gapless magnets with long-range magnetic order. Experimental investigations of the two-dimensional (2D) and three-dimensional (3D), gapless spin$-5/2$ (S$-5/2$), Heisenberg antiferromagnets (HAF) Rb$_2$MnF$_4$ \cite{Huberman} and MnF$_2$ \cite{Bayrakci} reveal symmetric Lorentzian lineshapes which broaden with temperature in excellent agreement with spin-wave calculations \cite{harris,Ty}. 
The basic assumption behind this description is that the spin-waves interact only weakly and the states available to them cover an extensive region of phase space. As the population of excited spin-waves increases with temperature, the rate of collisions between them also increases and they fluctuate within this large range of states in an uncorrelated manner. The damping is then due simply to loss of coherence associated with the reduced lifetime of the excitations.

The concept of thermal decoherence and the associated Lorentzian linewidth broadening have become so entrenched in current thinking that it is generally assumed to apply to all magnetic systems \cite{Forster,Marshall}. However for some magnets where there are strong interactions between the excitations and the phase space is restricted, it is not obvious that this reasoning should apply. Gapped antiferromagnets such as Haldane chains, spin ladders and dimer magnets which have a singlet ground state and triplet excitations are potential candidates.
For the S-1/2, dimer HAFs a dominant antiferromagnetic exchange interaction J$_0$, pairs the spins into dimers. The ground state is a singlet and the excited state is a triplet gapped by $\approx$J$_0$ while interdimer interactions allow the triplets to become dispersive and mobile. In the case of a large gap compared to the excitation bandwidth, the distribution of dimer excitations is severely limited. In addition, the dimers are subject to a hard-core constraint (only one excitation is allowed per dimer), which can be modeled as a strong, short-range repulsive interaction. As temperature is raised and the population of thermally excited dimers increases we would expect them to scatter in a highly correlated way due to the interactions between them and the limited density of states available to them. Furthermore, the thermal development of the lineshape of the excitations would presumably show evidence of these strong correlations.

Recently the temperature dependence of the magnetic excitation spectrum of copper nitrate (CN), a S$-1/2$ one-dimensional (1D) dimer HAF (alternating chain), has been investigated by means of inelastic neutron scattering (INS). This material is strongly dimerized and highly 1D which confines the excited states to a narrow band. The lineshape was found to change from a symmetric Lorentzian to a highly asymmetric profile with increasing temperature \cite{Tennant}. These results were attributed to strong correlations between the excitations and prove that far from becoming decoherent with temperature CN behaves as a 1D strongly correlated gas of hard-core Bosons. 1D systems often present a special case in magnetism where due to the geometry, collisions between the quasi-particles are unavoidable and interactions are often strong. It is important to determine whether a strongly correlated gas of hard-core Bosons is realizable in the more general case of a higher dimensional magnet where such collisions can be avoided.

Here, we investigate the temperature dependence of the magnons in the 3D  dimer HAF \scr. Although, the hard-core constraint also holds in this material, the 3D nature of the interdimer interactions along with a much smaller ratio of gap to bandwidth results in a much larger available phase space than in CN. In spite of the fact that \scr \, represents a much more generic magnetic system, we found that it also develops an asymmetric profile with increasing temperature. These results are important because they reveal that the conventional picture of thermal decoherence is invalid for a much wider range of systems than previously believed. A strongly correlated gas of magnetic quasi-particles can therefore be realized in a whole class of systems - dimerized magnets of all dimensionalities and maybe gapped magnets in general.

\begin{figure}[htb!]
	\centering
	\includegraphics[width=0.47\textwidth]{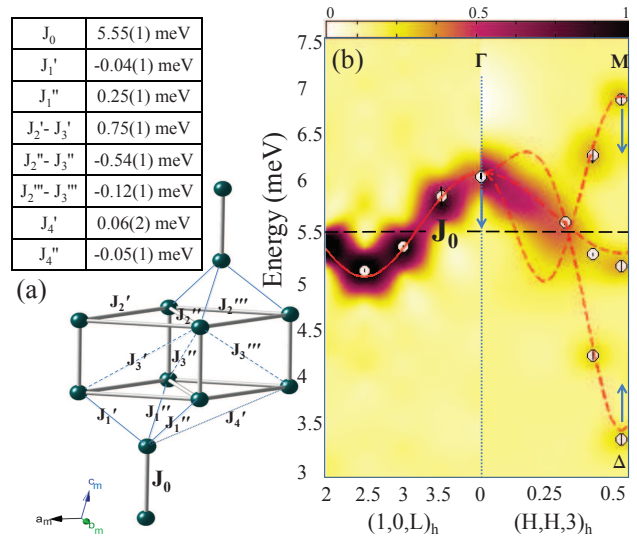}
        \caption{(Color online) (a) Low temperature (monoclinic) crystal structure of \scr \, showing the magnetic Cr$^{5+}$ ions only. The exchange interactions are labeled on the diagram and listed in the Table \cite{Quintero}.         
        (b) Constant-wavevector scans along $(1,0,L)_h$ and $(H,H,3)_h$ measured on FLEX at $1.6$\,K are interpolated to obtain a colormap where color indicates the relative scattered intensity. The white circles give the extracted peaks positions and the red lines correspond to the RPA model for the three monoclinic twins assuming the listed exchange constants. The blue arrows point in the direction of the temperature-induced energy shift, as explained in the text.}
    \label{fig:Figure1}
\end{figure}

\scr \, consists of S-1/2 Cr$^{5+}$ ions, which form hexagonal bilayers that are stacked in an ABCABC manner. The strong antiferromagnetic intrabilayer coupling, J$_0$, pairs the Cr$^{5+}$ ions into dimers giving rise to a singlet ground state [see Fig.\ \ref{fig:Figure1}(a)] \cite{Singh}. The gapped triplet excitations are dispersive in all crystallographic directions due to the 3D nature of the interdimer interactions. It was recently discovered that \scr \ undergoes a Jahn-Teller distortion that lifts the degeneracy of the singly occupied $e$-orbitals of Cr$^{5+}$ and lowers the crystal symmetry from hexagonal (R-3m) to monoclinic (C2/c) below $T_\text{JT}=285$\,K \cite{Chapon,WangESR}. Surprisingly THz and infrared spectroscopy (Ref.\ \onlinecite{WangESR}) and Raman scattering (Ref.\ \onlinecite{Dirk}) find that the phonons of the monoclinic structure do not fully develop until much lower temperatures of $\sim120$\,K. Previous single-crystal INS measurements at base temperature (1.6\,K) revealed a gapped band of excitations dispersing between $\Delta=3.4(3)$\,meV and $E_{upper}=7.10(5)$\,meV with bandwidth $B=E_{upper}-\Delta=3.7$\,meV \cite{Quintero}. For most wavevectors three triplet modes were observed arising from the three monoclinic twins that form below $T_\text{JT}$ [see Fig.\ \ref{fig:Figure1}(b)]. The data were successfully compared to a Random Phase Approximation (RPA) (red line in Fig.\ \ref{fig:Figure1}(b)) which was used to extract the exchange constants listed in Fig.\ \ref{fig:Figure1}(a) \cite{Quintero}. Heat capacity and magnetocaloric effect measurements of the Temperature-Field phase boundary around the lower critical field ($H_\text{c1}=30.4$\,T) where the magnons first condense into the ground state, reveal a critical exponent of $\nu=0.65(2)$ in excellent agreement to the value of $0.67$ predicted for the 3D Bose Einstein Condensation universality class \cite{AczelLANL}. This confirms the 3D nature of the interdimer couplings and shows that the exchange interactions are Heisenberg with negligible anisotropy.

Single crystal samples of \scr \, were grown using the floating zone technique \cite{Nazmul}. The temperature dependence of the magnetic excitations were investigated by INS using two cold triple axis spectrometers, V2-FLEX at Helmholtz-Zentrum Berlin (HZB) and TASP at Paul-Scherrer Institut (PSI) \cite{Semadeni2001}. Both instruments were optimized for good energy resolution. They were operated in W-configuration using a vertically-focusing Pyrolitic Graphite (PG) monochromator and horizontally-focusing PG analyzer along with a Beryllium filter. For FLEX the final wavevector was fixed to $k_f=1.2$\,\AA$^{-1}$ giving an energy resolution of $0.087$\,meV and two co-aligned single crystals ($7.74$\,g) were used. One crystal ($3.69$\,g) was used for the TASP experiment and the final wavevector was fixed to $k_f=1.1$\,\AA$^{-1}$ giving an energy resolution of $0.059$\,meV. 

\begin{figure}[htb!]
	\centering
		\includegraphics[width=0.47\textwidth]{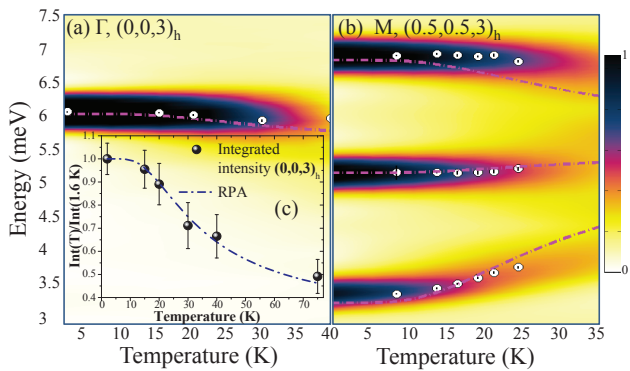}
				\caption {(Color online) Temperature-dependence of the constant-wavevector scans measured on FLEX at (a) $(0,0,3)_h$ and (b) $(0.5,0.5,3)_h$ compared to the RPA. The scans were interpolated and the color indicates the relative scattered intensity. The white circles correspond to the center of the peaks assuming a Lorentzian profile while the dashed magenta lines give the temperature evolution of the dispersion relation according to the RPA. (c) Integrated intensity of the excitation at $(0,0,3)_h$ as a function of temperature. The blue line gives the corresponding expectation values according to the RPA. }
	\label{fig:Figure2}
\end{figure}

The colorplots in Fig. \ref{fig:Figure2} show the energy scans performed on FLEX at $(0,0,3)_h$ and $(0.5,0.5,3)_h$ as a function of energy and temperature ($h$ indicates the high temperature hexagonal structure). $(0.5,0.5,3)_h$ corresponds to the $M$-point where the three crystallographic twins are inequivalent giving rise to three modes, while $(0,0,3)_h$ corresponds to the $\Gamma$-point where the three twins are degenerate. At both wavevectors, the dispersion renormalizes inwards towards the center of the band as the temperature is raised. Furthermore, the intensity of the scattering drops and the linewidth increases. The points on these colorplots indicate the centers of the peaks obtained by fitting the individual scans to Lorentzians. This function can only model the lineshape at low temperatures and is increasingly inaccurate at higher temperatures. The peaks become weaker with temperature and are difficult to detect above $T=75$\,K.

The lines in Fig. \ref{fig:Figure2} give the energy renormalization predicted by the RPA dimer model using the exchange constants listed in Fig. \ref{fig:Figure1}. The model follows the center of the excitations up to $40$\,K for $(0,0,3)_h$ and $25$\,K for the lowest and middle modes at $(0.5,0.5,3)_h$. The agreement with the upper mode is poor. The inwards renormalization of the dispersion occurs because as temperature is increased the dimers become thermally excited which interferes with the intersite interactions, thus the dispersion tends towards the intradimer exchange energy J$_0$.  Such a renormalization has been reported previously for other gapped dimer antiferromagnets \cite{Zheludev2008}, such as: BaCuSi$_2$O$_6$ \cite{SasagoPRB55}, KCuCl$_3$ \cite{Cavadini_T} and Pr$_3$Tl \cite{Houmann}. The strong decrease in integrated intensity of the mode at $(0,0,3)_h$ is shown in Fig.\ \ref{fig:Figure2}(c), which also agrees well with the predictions of RPA.

\begin{figure}[htb!]
	\centering
        \includegraphics[width=0.5\textwidth]{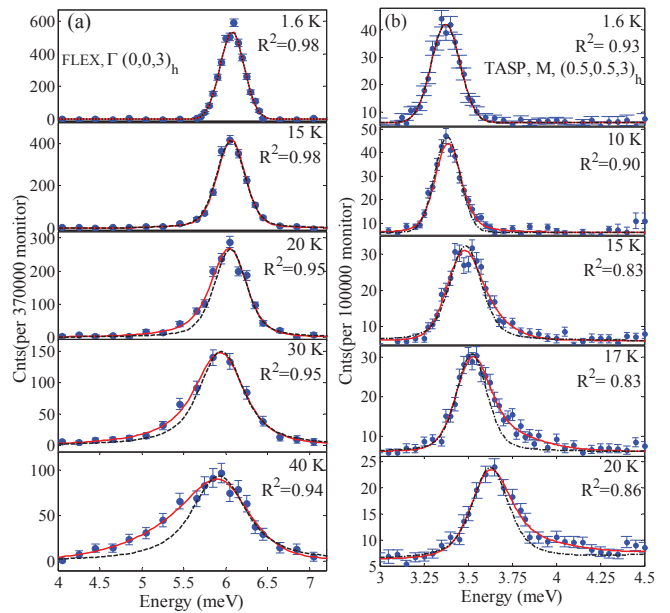} 		
				\caption {(Color online)	Constant-wavevector scans at different temperatures measured on (a) FLEX at $(0,0,3)_h$, and (b) TASP at $(0.5,0.5,3)_h$ compared to symmetric and asymmetric lineshape functions. The red lines correspond to the fit using Eq.\ \ref{eq:lineshape}, where the asymmetry parameters $\alpha$ and $\beta$ are varied. The black dashed lines correspond to the symmetric fit where $\alpha =\beta =0$. For the FLEX data the background was subtracted for clarity and the Gaussian width is fixed to $\sigma_{FLEX}= 0.143$\,meV. For TASP, the Gaussian width is fixed to $\sigma_{TASP}= 0.079$\,meV.}
	\label{fig:Figure3}
\end{figure}

The individual scans at $(0,0,3)_h$ and $(0.5,0.5,3)_h$ are displayed for different temperatures in Fig.\ \ref{fig:Figure3}. From these plots it is immediately clear that the lineshapes of the excitations do not broaden symmetrically according to standard expectations, but become increasingly asymmetric as temperature is raised. At $(0,0,3)_h$  where the three twins are degenerate, the mode is at 6\,meV which is close to the upper edge of the excitation bandwidth [Fig.\ \ref{fig:Figure3}(a)]. As temperature is increased the peak becomes asymmetric with a tail extending towards lower energies. At $(0.5,0.5,3)_h$ the lowest mode which lies at the bottom of the band again reveals temperature-induced asymmetry [Fig.\ \ref{fig:Figure3}(b)] but now the tail extends towards higher energies. These results are an important departure from existing theory that predicts a Lorentzian profile which broadens symmetrically due to thermal decoherence \cite{Bayrakci, Cavadini_T, Damle, Huberman, Xu}. Lineshapes can appear asymmetric due to an artifact of the instrumental resolution, however our lowest temperature peaks are symmetric showing that the resolution which is temperature independent is good enough to prevent this effect. A simulation of the RPA dispersion convoluted with the instrumental resolution calculated using the program \textsc{Rescal} (Ref.\ \onlinecite{rescal}) confirms that a symmetric Voigt lineshape should be observed unless there is intrinsic asymmetry.

To parameterize the temperature evolution of the lineshape, the individual scans acquired on FLEX and TASP were fitted to an asymmetric peak function \cite{Tennant}. It is based on a Lorentzian but where the usual argument in the denominator $[1+t^{2}]$ is replaced by the polynomial $[1+(t-\alpha t^{2} +\beta t^{3})]$. The Lorentzian is then convolved with the experimental resolution assumed to be Gaussian:
\begin{eqnarray}
I(\vec{Q}, \omega)= A(Q)\int^{\infty}_{-\infty}dt \frac{exp[-\frac{(\omega-\gamma(Q)t-d(Q))^2}{2\sigma^2}]}{\sqrt{2\pi\sigma^2}}
\frac{1}{\pi} \nonumber \\
\times \frac{1}{1+(t-\alpha(Q)t^2+\beta(Q)t^3)^2}.
\label{eq:lineshape}
\end{eqnarray}
In this function, $\sigma$ is the variance of the Gaussian, $d(Q)$ is the center of the peak and $A(Q)$ is the overall amplitude. $\gamma$ is the Lorentzian width and the asymmetry is controlled by $\alpha(Q)t^2$ and a further damping term $\beta(Q)t^3$. The sign of $\alpha(Q)t^2$ governs whether the peak broadens to higher or lower energies. 
\begin{figure}[htb!]
	\centering		
	\includegraphics[width=0.48\textwidth]{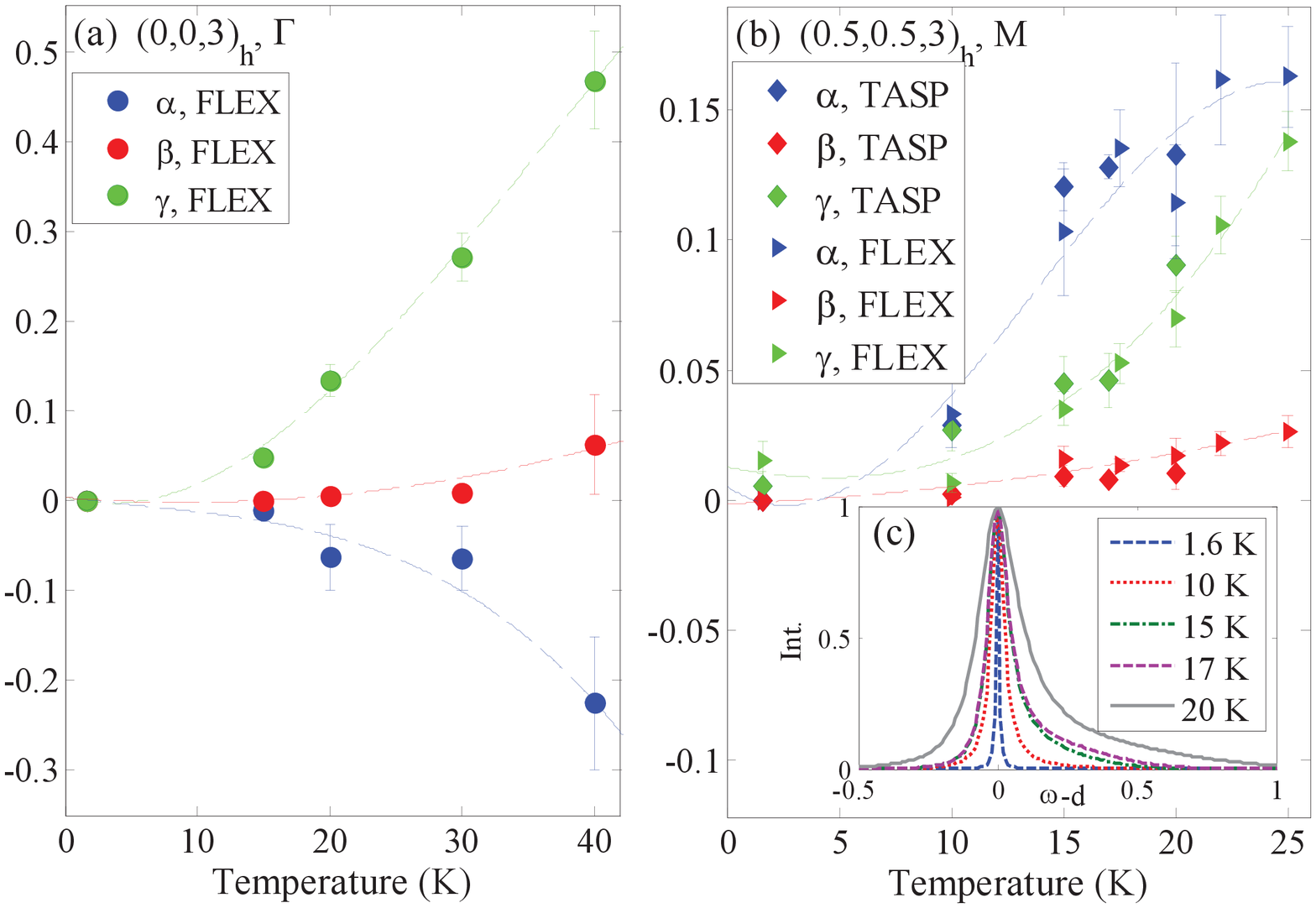} 
				\caption { (Color online) 
				Lorentzian width $\gamma$ and asymmetry terms $\alpha$ and $\beta$ extracted from fits of Eq.\ \ref{eq:lineshape} to (a) the unique peak at $(0,0,3)_h$ measured on FLEX and (b) the lowest energy mode at $(0.5,0.5,3)_h$ from FLEX and TASP. The dashed lines are guides to the eye. (c) The inset shows the peak profiles from the fit of the TASP data de-convolved from the resolution function.}
	\label{fig:Figure4}
\end{figure} 

This function was used to fit the data. The lowest temperature peaks (1.6\,K) which are symmetric were fitted by fixing $\alpha$ and $\beta$ to zero, to give a symmetric Voigt profile. The excitations have two broadening mechanisms. Firstly due to the instrumental resolution and mosaic spread of the sample which give a Gaussian profile, and secondly due to the finite lifetime of the magnons which produce a Lorentzian profile. The resolution and mosaic are temperature-independent, therefore the Gaussian width was fixed to the base temperature value. Above base temperature $\alpha$ and $\beta$ were allowed to vary and the best fits are given by the solid red lines in Fig.\ \ref{fig:Figure3} while the fitted parameters are displayed in Fig.\ \ref{fig:Figure4}. The Lorentzian width $\gamma$ increases rapidly with temperature while $\beta$ increases slowly. The asymmetry term $\alpha$ becomes negative with increasing temperature for the excitation at $(0,0,3)_h$ indicating that the asymmetry is weighted towards lower energies, while it is positive for the low energy mode at $(0.5,0.5,3)_h$ indicating that here the asymmetry is towards higher energies. From these results it is clear that the asymmetry tends towards the center of the band [see arrows in Fig.\ \ref{fig:Figure1}(b)]. For comparison symmetric lineshapes were also fitted to the data by setting $\alpha$ and $\beta$ to zero for all temperatures as shown by the black lines in Fig.\ \ref{fig:Figure3}. It is clear that the symmetric profile is completely unable to model the data at and above 20\,K at $(0,0,3)_h$, and at and above 15\,K at $(0.5,0.5,3)_h$.

In summary, we have investigated the temperature dependence of the excitations in the 3D gapped dimer HAF \scr \ using high resolution INS. The results reveal an inwards renormalization of the dispersion towards the intradimer exchange constant (J$_0$) with increasing temperature. In addition, the modes are damped and their intensity decreases as temperature is raised. Most importantly the lineshapes of the excitations become asymmetric with temperature being weighted towards the center of the band, quite unlike the symmetric Lorentzian lineshape expected from spin-wave theory. The RPA is able to model the reduction in the integrated intensity and renormalization for temperatures below 25\,K. On the other hand the model described here predicts sharp excitations at all temperatures because it ignores correlations between the magnons, RPA can be modified to include damping effects \cite{Bak, Cavadini_T, Jensen_2010}. However, it then predicts that, the excitations would broaden symmetrically with a Lorentzian lineshape for temperatures $T<\Delta/k_\text{B} =40$\,K. This contrasts strongly with our experimental observation that the lineshape becomes asymmetric for temperatures above $15$\,K. 

Typically the lineshapes of magnetic excitations are believed to broaden symmetrically with a Lorentzian profile as temperature increases. The mechanism is assumed to be a reduction in lifetime due to thermal decoherence as the excitations  scattering off each other \cite{Bayrakci, Cavadini_T, Damle, Huberman, Xu}. The asymmetry observed in \scr \ is therefore unprecedented because it suggests that far from becoming increasingly incoherent, the magnons interact strongly with each other and scatter in a coherent manner. The dimer excitations interact because of the interdimer exchange constants and the hard-core constraint which can be viewed as a strong short-range repulsion. When a magnon is created at finite temperatures by the neutron scattering process, it will interact and scatter from the thermally excited magnons. The energy required to create it at finite temperature is then different from that required at zero temperature because the energy of interaction must also be considered. As a result the observed dispersion relation and lineshape are broadened predominantly in the direction of the highest density of states. Rather than creating individual dimer excitations, their correlations within the gas of thermally excited dimers must also be taken into account.

There is to our knowledge only one conclusive prior observation of asymmetric thermal lineshape broadening in the alternating chain compound CN \cite{Tennant}. This material is highly 1D and strongly dimerized with a very large ratio of gap ($\Delta=0.35$\,meV) to excitation bandwidth ($\Delta / B = 3.5$). The asymmetry is attributed to the hard-core constraint and the severely restricted phase space due to the large gap and one-dimensionality. A theoretical model was developed applicable to this special case which agreed well with the data \cite{Tennant, Essler2008, James, James_Thesis}. The asymmetry was also parameterized as a function of temperature using Eq.\ \ref{eq:lineshape} as described in Ref.\ \onlinecite{Tennant} and can be compared to \scr. After rescaling the \scr\ data by the CN intradimer exchange constant and refitting, the values for the Lorentzian width $\gamma$ and asymmetry $\alpha$ were found to be similar (within 30\%) to those of CN. 

The dimers in \scr \ are coupled in 3D and the gap is much smaller than for CN. The ratio of gap to bandwidth is $0.9$ in \scr \, compared to $3.5$ for CN, therefore, the phase space available for scattering should be much larger and the interactions weaker. Therefore, the phase space available for scattering should be much larger and the interactions between magnons weaker. In spite of this, the asymmetry of the lineshape is still clearly evident and of similar magnitude.  

These important results contradict the long-standing view that excitations become increasingly incoherent and short-lived as temperature increases. Instead, they suggest that a strongly correlated gas of magnetic excitations is not confined to the special case of highly-dimerized, 1D antiferromagnets, but is realized much more generally in dimerized HAFs of all dimensionalities and gap sizes and perhaps in gapped magnets generally.\\

We thank F. H. L. Essler, A. J. A. James, D. A. Tennant, F. Groitl, K. Habicht and A. M. Tsvelik for helpful discussions. This work is partially based on experiments performed at the Swiss spallation neutron source SINQ, Paul Scherrer Institute, Villigen, Switzerland.


\begin{thebibliography}{27}
\expandafter\ifx\csname natexlab\endcsname\relax\def\natexlab#1{#1}\fi
\expandafter\ifx\csname bibnamefont\endcsname\relax
  \def\bibnamefont#1{#1}\fi
\expandafter\ifx\csname bibfnamefont\endcsname\relax
  \def\bibfnamefont#1{#1}\fi
\expandafter\ifx\csname citenamefont\endcsname\relax
  \def\citenamefont#1{#1}\fi
\expandafter\ifx\csname url\endcsname\relax
  \def\url#1{\texttt{#1}}\fi
\expandafter\ifx\csname urlprefix\endcsname\relax\def\urlprefix{URL }\fi
\providecommand{\bibinfo}[2]{#2}
\providecommand{\eprint}[2][]{\url{#2}}

\bibitem[{\citenamefont{Harris et~al.}(1971)\citenamefont{Harris, Kumar,
  Halperin, and Hohenberg}}]{harris}
\bibinfo{author}{\bibfnamefont{A.~B.} \bibnamefont{Harris}},
  \bibinfo{author}{\bibfnamefont{D.}~\bibnamefont{Kumar}},
  \bibinfo{author}{\bibfnamefont{B.~I.} \bibnamefont{Halperin}},
  \bibnamefont{and} \bibinfo{author}{\bibfnamefont{P.~C.}
  \bibnamefont{Hohenberg}}, \bibinfo{journal}{Phys. Rev. B}
  \textbf{\bibinfo{volume}{3}}, \bibinfo{pages}{961} (\bibinfo{year}{1971}).

\bibitem[{\citenamefont{Tyc and Halperin}(1990)}]{Ty}
\bibinfo{author}{\bibfnamefont{S.}~\bibnamefont{Tyc}} \bibnamefont{and}
  \bibinfo{author}{\bibfnamefont{B.~I.} \bibnamefont{Halperin}},
  \bibinfo{journal}{Phys. Rev. B} \textbf{\bibinfo{volume}{42}},
  \bibinfo{pages}{2096} (\bibinfo{year}{1990}).

\bibitem[{\citenamefont{Huberman et~al.}(2008)\citenamefont{Huberman, Tennant,
  Cowley, Coldea, and Forst}}]{Huberman}
\bibinfo{author}{\bibfnamefont{T.}~\bibnamefont{Huberman}},
  \bibinfo{author}{\bibfnamefont{D.~A.} \bibnamefont{Tennant}},
  \bibinfo{author}{\bibfnamefont{R.~A.} \bibnamefont{Cowley}},
  \bibinfo{author}{\bibfnamefont{R.}~\bibnamefont{Coldea}}, \bibnamefont{and}
  \bibinfo{author}{\bibfnamefont{C.~D.} \bibnamefont{Forst}},
  \bibinfo{journal}{J. Stat. Mech.} p. \bibinfo{pages}{05017}
  (\bibinfo{year}{2008}).

\bibitem[{\citenamefont{Bayrakci et~al.}(2006)\citenamefont{Bayrakci, Keller,
  Habicht, and Keimer}}]{Bayrakci}
\bibinfo{author}{\bibfnamefont{S.~P.} \bibnamefont{Bayrakci}},
  \bibinfo{author}{\bibfnamefont{T.}~\bibnamefont{Keller}},
  \bibinfo{author}{\bibfnamefont{K.}~\bibnamefont{Habicht}}, \bibnamefont{and}
  \bibinfo{author}{\bibfnamefont{B.}~\bibnamefont{Keimer}},
  \bibinfo{journal}{Science} \textbf{\bibinfo{volume}{312}},
  \bibinfo{pages}{5782} (\bibinfo{year}{2006}).

\bibitem[{\citenamefont{Forster}(1975)}]{Forster}
\bibinfo{author}{\bibfnamefont{D.}~\bibnamefont{Forster}},
  \emph{\bibinfo{title}{Hydrodynamic Fluctuations, Broken Symmetry, and
  Correlation Functions}} (\bibinfo{publisher}{W.A. Benjamin, Reading},
  \bibinfo{year}{1975}).

\bibitem[{\citenamefont{Marshall and Lovesey}(1971)}]{Marshall}
\bibinfo{author}{\bibfnamefont{W.}~\bibnamefont{Marshall}} \bibnamefont{and}
  \bibinfo{author}{\bibfnamefont{S.~W.} \bibnamefont{Lovesey}},
  \emph{\bibinfo{title}{Theory of thermal neutron scattering}}
  (\bibinfo{publisher}{Oxford University Press}, \bibinfo{year}{1971}).

\bibitem[{\citenamefont{Tennant et~al.}(2012)\citenamefont{Tennant, Lake,
  James, Essler, Notbohm, Mikeska, Fielden, K\"ogerler, Canfield, and
  Telling}}]{Tennant}
\bibinfo{author}{\bibfnamefont{D.~A.} \bibnamefont{Tennant}},
  \bibinfo{author}{\bibfnamefont{B.}~\bibnamefont{Lake}},
  \bibinfo{author}{\bibfnamefont{A.~J.~A.} \bibnamefont{James}},
  \bibinfo{author}{\bibfnamefont{F.~H.~L.} \bibnamefont{Essler}},
  \bibinfo{author}{\bibfnamefont{S.}~\bibnamefont{Notbohm}},
  \bibinfo{author}{\bibfnamefont{H.-J.} \bibnamefont{Mikeska}},
  \bibinfo{author}{\bibfnamefont{J.}~\bibnamefont{Fielden}},
  \bibinfo{author}{\bibfnamefont{P.}~\bibnamefont{K\"ogerler}},
  \bibinfo{author}{\bibfnamefont{P.~C.} \bibnamefont{Canfield}},
  \bibnamefont{and} \bibinfo{author}{\bibfnamefont{M.~T.~F.}
  \bibnamefont{Telling}}, \bibinfo{journal}{Phys. Rev. B}
  \textbf{\bibinfo{volume}{85}}, \bibinfo{pages}{014402}
  (\bibinfo{year}{2012}).

\bibitem[{\citenamefont{Singh and Johnston}(2007)}]{Singh}
\bibinfo{author}{\bibfnamefont{Y.}~\bibnamefont{Singh}} \bibnamefont{and}
  \bibinfo{author}{\bibfnamefont{D.~C.} \bibnamefont{Johnston}},
  \bibinfo{journal}{Phys. Rev. B} \textbf{\bibinfo{volume}{76}},
  \bibinfo{pages}{012407} (\bibinfo{year}{2007}).

\bibitem[{\citenamefont{Chapon et~al.}(2008)\citenamefont{Chapon, Stock,
  Radaelli, and Martin}}]{Chapon}
\bibinfo{author}{\bibfnamefont{L.~C.} \bibnamefont{Chapon}},
  \bibinfo{author}{\bibfnamefont{C.}~\bibnamefont{Stock}},
  \bibinfo{author}{\bibfnamefont{P.~G.} \bibnamefont{Radaelli}},
  \bibnamefont{and} \bibinfo{author}{\bibfnamefont{C.}~\bibnamefont{Martin}},
  \bibinfo{journal}{arXiv:0807.0877v2 [cond-mat.mtrl-sci]}
  (\bibinfo{year}{2008}).

\bibitem[{\citenamefont{Wang et~al.}(2011)\citenamefont{Wang, Schmidt, Gunther,
  Schaile, Pascher, Mayr, Goncharov, Quintero-Castro, Islam, Lake
  et~al.}}]{WangESR}
\bibinfo{author}{\bibfnamefont{Z.}~\bibnamefont{Wang}},
  \bibinfo{author}{\bibfnamefont{M.}~\bibnamefont{Schmidt}},
  \bibinfo{author}{\bibfnamefont{A.}~\bibnamefont{Gunther}},
  \bibinfo{author}{\bibfnamefont{S.}~\bibnamefont{Schaile}},
  \bibinfo{author}{\bibfnamefont{N.}~\bibnamefont{Pascher}},
  \bibinfo{author}{\bibfnamefont{F.}~\bibnamefont{Mayr}},
  \bibinfo{author}{\bibfnamefont{Y.}~\bibnamefont{Goncharov}},
  \bibinfo{author}{\bibfnamefont{D.~L.} \bibnamefont{Quintero-Castro}},
  \bibinfo{author}{\bibfnamefont{A.~T. M.~N.} \bibnamefont{Islam}},
  \bibinfo{author}{\bibfnamefont{B.}~\bibnamefont{Lake}}, \bibnamefont{et~al.},
  \bibinfo{journal}{Phys. Rev. B} \textbf{\bibinfo{volume}{83}},
  \bibinfo{pages}{201102(R)} (\bibinfo{year}{2011}).

\bibitem[{\citenamefont{Wulferding et~al.}(2011)\citenamefont{Wulferding,
  Lemmens, Choi, Gnezdilov, Pashkevich, Deisenhofer, Quintero-Castro,
  NazmulIslam, and Lake}}]{Dirk}
\bibinfo{author}{\bibfnamefont{D.}~\bibnamefont{Wulferding}},
  \bibinfo{author}{\bibfnamefont{P.}~\bibnamefont{Lemmens}},
  \bibinfo{author}{\bibfnamefont{K.~Y.} \bibnamefont{Choi}},
  \bibinfo{author}{\bibfnamefont{V.}~\bibnamefont{Gnezdilov}},
  \bibinfo{author}{\bibfnamefont{Y.~G.} \bibnamefont{Pashkevich}},
  \bibinfo{author}{\bibfnamefont{J.}~\bibnamefont{Deisenhofer}},
  \bibinfo{author}{\bibfnamefont{D.}~\bibnamefont{Quintero-Castro}},
  \bibinfo{author}{\bibfnamefont{A.~T.~M.} \bibnamefont{NazmulIslam}},
  \bibnamefont{and} \bibinfo{author}{\bibfnamefont{B.}~\bibnamefont{Lake}},
  \bibinfo{journal}{Phys. Rev. B} \textbf{\bibinfo{volume}{84}},
  \bibinfo{pages}{064419} (\bibinfo{year}{2011}).

\bibitem[{\citenamefont{Quintero-Castro
  et~al.}(2010)\citenamefont{Quintero-Castro, Lake, Wheeler, Islam, Guidi,
  Rule, Izaola, Russina, Kiefer, and Skoursky}}]{Quintero}
\bibinfo{author}{\bibfnamefont{D.~L.} \bibnamefont{Quintero-Castro}},
  \bibinfo{author}{\bibfnamefont{B.}~\bibnamefont{Lake}},
  \bibinfo{author}{\bibfnamefont{E.~M.} \bibnamefont{Wheeler}},
  \bibinfo{author}{\bibfnamefont{A.~T. M.~N.} \bibnamefont{Islam}},
  \bibinfo{author}{\bibfnamefont{T.}~\bibnamefont{Guidi}},
  \bibinfo{author}{\bibfnamefont{K.~C.} \bibnamefont{Rule}},
  \bibinfo{author}{\bibfnamefont{Z.}~\bibnamefont{Izaola}},
  \bibinfo{author}{\bibfnamefont{M.}~\bibnamefont{Russina}},
  \bibinfo{author}{\bibfnamefont{K.}~\bibnamefont{Kiefer}}, \bibnamefont{and}
  \bibinfo{author}{\bibfnamefont{Y.}~\bibnamefont{Skoursky}},
  \bibinfo{journal}{Phys. Rev. B} \textbf{\bibinfo{volume}{81}},
  \bibinfo{pages}{014415} (\bibinfo{year}{2010}).

\bibitem[{\citenamefont{Aczel et~al.}(2009)\citenamefont{Aczel, Kohama,
  Marcenat, Weickert, Jaime, Ayala-Valenzuela, McDonald, Selesnic, Dabkowska,
  and Luke}}]{AczelLANL}
\bibinfo{author}{\bibfnamefont{A.~A.} \bibnamefont{Aczel}},
  \bibinfo{author}{\bibfnamefont{Y.}~\bibnamefont{Kohama}},
  \bibinfo{author}{\bibfnamefont{C.}~\bibnamefont{Marcenat}},
  \bibinfo{author}{\bibfnamefont{F.}~\bibnamefont{Weickert}},
  \bibinfo{author}{\bibfnamefont{M.}~\bibnamefont{Jaime}},
  \bibinfo{author}{\bibfnamefont{O.~E.} \bibnamefont{Ayala-Valenzuela}},
  \bibinfo{author}{\bibfnamefont{R.~D.} \bibnamefont{McDonald}},
  \bibinfo{author}{\bibfnamefont{S.~D.} \bibnamefont{Selesnic}},
  \bibinfo{author}{\bibfnamefont{H.~A.} \bibnamefont{Dabkowska}},
  \bibnamefont{and} \bibinfo{author}{\bibfnamefont{G.~M.} \bibnamefont{Luke}},
  \bibinfo{journal}{Phys. Rev. Lett.} \textbf{\bibinfo{volume}{103}},
  \bibinfo{pages}{207203} (\bibinfo{year}{2009}).

\bibitem[{\citenamefont{Islam et~al.}(2010)\citenamefont{Islam,
  Quintero-Castro, Lake, Siemensmeyer, Kiefer, Skourski, and
  Herrmannsdorfer}}]{Nazmul}
\bibinfo{author}{\bibfnamefont{A.~T. M.~N.} \bibnamefont{Islam}},
  \bibinfo{author}{\bibfnamefont{D.~L.} \bibnamefont{Quintero-Castro}},
  \bibinfo{author}{\bibfnamefont{B.}~\bibnamefont{Lake}},
  \bibinfo{author}{\bibfnamefont{K.}~\bibnamefont{Siemensmeyer}},
  \bibinfo{author}{\bibfnamefont{K.}~\bibnamefont{Kiefer}},
  \bibinfo{author}{\bibfnamefont{Y.}~\bibnamefont{Skourski}}, \bibnamefont{and}
  \bibinfo{author}{\bibfnamefont{T.}~\bibnamefont{Herrmannsdorfer}},
  \bibinfo{journal}{Crys. Growth \& Des.} \textbf{\bibinfo{volume}{10}},
  \bibinfo{pages}{465} (\bibinfo{year}{2010}).

\bibitem[{\citenamefont{Semadeni et~al.}(2001)\citenamefont{Semadeni, Roessli,
  and B\"oni}}]{Semadeni2001}
\bibinfo{author}{\bibfnamefont{F.}~\bibnamefont{Semadeni}},
  \bibinfo{author}{\bibfnamefont{B.}~\bibnamefont{Roessli}}, \bibnamefont{and}
  \bibinfo{author}{\bibfnamefont{P.}~\bibnamefont{B\"oni}},
  \bibinfo{journal}{Physica B.} \textbf{\bibinfo{volume}{297}},
  \bibinfo{pages}{152} (\bibinfo{year}{2001}).

\bibitem[{\citenamefont{Zheludev et~al.}(2008)\citenamefont{Zheludev, Garlea,
  Regnault, Manaka, Tsvelik, and Chung}}]{Zheludev2008}
\bibinfo{author}{\bibfnamefont{A.}~\bibnamefont{Zheludev}},
  \bibinfo{author}{\bibfnamefont{V.~O.} \bibnamefont{Garlea}},
  \bibinfo{author}{\bibfnamefont{L.-P.} \bibnamefont{Regnault}},
  \bibinfo{author}{\bibfnamefont{H.}~\bibnamefont{Manaka}},
  \bibinfo{author}{\bibfnamefont{A.}~\bibnamefont{Tsvelik}}, \bibnamefont{and}
  \bibinfo{author}{\bibfnamefont{J.-H.} \bibnamefont{Chung}},
  \bibinfo{journal}{Phys. Rev. Lett.} \textbf{\bibinfo{volume}{100}},
  \bibinfo{pages}{157204} (\bibinfo{year}{2008}).

\bibitem[{\citenamefont{Sasago et~al.}(1997)\citenamefont{Sasago, Uchinokura,
  Zheludev, and Shirane}}]{SasagoPRB55}
\bibinfo{author}{\bibfnamefont{Y.}~\bibnamefont{Sasago}},
  \bibinfo{author}{\bibfnamefont{K.}~\bibnamefont{Uchinokura}},
  \bibinfo{author}{\bibfnamefont{A.}~\bibnamefont{Zheludev}}, \bibnamefont{and}
  \bibinfo{author}{\bibfnamefont{G.}~\bibnamefont{Shirane}},
  \bibinfo{journal}{Phys. Rev. B} \textbf{\bibinfo{volume}{55}},
  \bibinfo{pages}{8357} (\bibinfo{year}{1997}).

\bibitem[{\citenamefont{Cavadini et~al.}(2000)\citenamefont{Cavadini, R\"uegg,
  Henggeler, Furrer, G\"udel, Kr\"amer, and Mutka}}]{Cavadini_T}
\bibinfo{author}{\bibfnamefont{N.}~\bibnamefont{Cavadini}},
  \bibinfo{author}{\bibfnamefont{C.}~\bibnamefont{R\"uegg}},
  \bibinfo{author}{\bibfnamefont{W.}~\bibnamefont{Henggeler}},
  \bibinfo{author}{\bibfnamefont{A.}~\bibnamefont{Furrer}},
  \bibinfo{author}{\bibfnamefont{H.~U.} \bibnamefont{G\"udel}},
  \bibinfo{author}{\bibfnamefont{K.}~\bibnamefont{Kr\"amer}}, \bibnamefont{and}
  \bibinfo{author}{\bibfnamefont{H.}~\bibnamefont{Mutka}},
  \bibinfo{journal}{Eur. Phys. J. B} \textbf{\bibinfo{volume}{18}},
  \bibinfo{pages}{565} (\bibinfo{year}{2000}).

\bibitem[{\citenamefont{Houmann et~al.}(1975)\citenamefont{Houmann, Chapellier,
  Mackintosh, Bak, McMasters, and Gschneidner}}]{Houmann}
\bibinfo{author}{\bibfnamefont{J.~G.} \bibnamefont{Houmann}},
  \bibinfo{author}{\bibfnamefont{M.}~\bibnamefont{Chapellier}},
  \bibinfo{author}{\bibfnamefont{A.~R.} \bibnamefont{Mackintosh}},
  \bibinfo{author}{\bibfnamefont{P.}~\bibnamefont{Bak}},
  \bibinfo{author}{\bibfnamefont{O.~D.} \bibnamefont{McMasters}},
  \bibnamefont{and} \bibinfo{author}{\bibfnamefont{K.~A.}
  \bibnamefont{Gschneidner}}, \bibinfo{journal}{Phys. Rev. Lett.}
  \textbf{\bibinfo{volume}{34}}, \bibinfo{pages}{587} (\bibinfo{year}{1975}).

\bibitem[{\citenamefont{Damle and Sachdev}(1998)}]{Damle}
\bibinfo{author}{\bibfnamefont{K.}~\bibnamefont{Damle}} \bibnamefont{and}
  \bibinfo{author}{\bibfnamefont{S.}~\bibnamefont{Sachdev}},
  \bibinfo{journal}{Phy. Rev. B.} \textbf{\bibinfo{volume}{57}},
  \bibinfo{pages}{8307} (\bibinfo{year}{1998}).

\bibitem[{\citenamefont{Xu et~al.}(2007)\citenamefont{Xu, Broholm, Soh, Aeppli,
  Di~Tusa, Chen, Kenzelmann, Frost, Ito, Oka et~al.}}]{Xu}
\bibinfo{author}{\bibfnamefont{G.}~\bibnamefont{Xu}},
  \bibinfo{author}{\bibfnamefont{C.}~\bibnamefont{Broholm}},
  \bibinfo{author}{\bibfnamefont{Y.}~\bibnamefont{Soh}},
  \bibinfo{author}{\bibfnamefont{G.}~\bibnamefont{Aeppli}},
  \bibinfo{author}{\bibfnamefont{J.~F.} \bibnamefont{Di~Tusa}},
  \bibinfo{author}{\bibfnamefont{Y.}~\bibnamefont{Chen}},
  \bibinfo{author}{\bibfnamefont{M.}~\bibnamefont{Kenzelmann}},
  \bibinfo{author}{\bibfnamefont{C.~D.} \bibnamefont{Frost}},
  \bibinfo{author}{\bibfnamefont{T.}~\bibnamefont{Ito}},
  \bibinfo{author}{\bibfnamefont{K.}~\bibnamefont{Oka}}, \bibnamefont{et~al.},
  \bibinfo{journal}{Science} \textbf{\bibinfo{volume}{317}},
  \bibinfo{pages}{1049} (\bibinfo{year}{2007}).

\bibitem[{\citenamefont{http://www.ill.eu/instruments-support/computing-for-science/cs-software/all-software/matlab-ill/rescal-for
  matlab/}()}]{rescal}
\bibinfo{author}{\bibnamefont{http://www.ill.eu/instruments-support/computing-for-science/cs-software/all-software/matlab-ill/rescal-for
  matlab/}}.

\bibitem[{\citenamefont{Bak}(1975)}]{Bak}
\bibinfo{author}{\bibfnamefont{P.}~\bibnamefont{Bak}}, \bibinfo{journal}{Phy.
  Rev.} \textbf{\bibinfo{volume}{12}}, \bibinfo{pages}{5203}
  (\bibinfo{year}{1975}).

\bibitem[{\citenamefont{Jensen}(2011)}]{Jensen_2010}
\bibinfo{author}{\bibfnamefont{J.}~\bibnamefont{Jensen}},
  \bibinfo{journal}{Phys. Rev. B} \textbf{\bibinfo{volume}{83}},
  \bibinfo{pages}{064420} (\bibinfo{year}{2011}).

\bibitem[{\citenamefont{Essler and Konik}(2008)}]{Essler2008}
\bibinfo{author}{\bibfnamefont{F.~H.~L.} \bibnamefont{Essler}}
  \bibnamefont{and} \bibinfo{author}{\bibfnamefont{R.~M.} \bibnamefont{Konik}},
  \bibinfo{journal}{Phy. Rev. B} \textbf{\bibinfo{volume}{78}},
  \bibinfo{pages}{100403} (\bibinfo{year}{2008}).

\bibitem[{\citenamefont{James et~al.}(2008)\citenamefont{James, Essler, and
  Konik}}]{James}
\bibinfo{author}{\bibfnamefont{A.~J.~A.} \bibnamefont{James}},
  \bibinfo{author}{\bibfnamefont{F.~H.~L.} \bibnamefont{Essler}},
  \bibnamefont{and} \bibinfo{author}{\bibfnamefont{R.~M.} \bibnamefont{Konik}},
  \bibinfo{journal}{Phys. Rev. B} \textbf{\bibinfo{volume}{78}},
  \bibinfo{pages}{094411} (\bibinfo{year}{2008}).

\bibitem[{\citenamefont{James}(2008)}]{James_Thesis}
\bibinfo{author}{\bibfnamefont{A.~J.~A.} \bibnamefont{James}}, Ph.D. thesis,
  \bibinfo{school}{University of Oxford} (\bibinfo{year}{2008}).

\end{thebibliography}

\end{document}